\documentclass{PoS}
\usepackage{amsmath,epsfig,axodraw}
\title{Two-loop results on the renormalization of vacuum expectation
  values and infrared divergences in the FDH scheme}

\ShortTitle{Two-loop results on the renormalization of vevs and
  infrared divergences in FDH}

\author{Christoph Gnendiger\\
      Institut f\"ur Kern- und Teilchenphysik, Germany}
\author{Adrian Signer\\
       Paul Scherrer Institut,
CH-5232 Villigen PSI, Switzerland and\\ Physik-Institut, Universität Zürich, Switzerland
}
\author{Marcus Sperling\\
        Institut f\"ur Theoretische Physik, Universit\"at Hannover, Germany}
\author{\speaker{Dominik St\"ockinger}%
\\
        Institut f\"ur Kern- und Teilchenphysik, TU Dresden, Germany\\
        E-mail: \email{Dominik.Stoeckinger@tu-dresden.de}}
\author{Alexander Voigt\\
        Institut f\"ur Kern- und Teilchenphysik, TU Dresden, Germany}

\abstract{Recent progress in the understanding of vacuum
  expectation values and of infrared divergences in different
  regularization schemes is reviewed. Vacuum
  expectation values are gauge and renormalization-scheme dependent
  quantities. Using a method based on Slavnov-Taylor identities, the
  renormalization properties could be better understood. The practical
outcome is the computation of the $\beta$ functions for vacuum
expectation values in general gauge theories. 

The infrared structure
of gauge theory amplitudes depends on the regularization scheme. The
well-known prediction of the infrared structure in CDR can be
generalized to the FDH and DRED schemes and is in agreement with
explicit computations of the quark and gluon form factors. We discuss
particularly the correct renormalization procedure and the distinction
between \MSbar\ and \DRbar\ renormalization. An important practical
outcome are transition rules between CDR and FDH amplitudes.}

\FullConference{ Loops and Legs in Quantum Field Theory - LL 2014,\\
		 27 April - 2 May 2014 \\
		 Weimar, Germany }

\newcommand{\MSbar}{\ensuremath{\overline{\text{MS}}}}
\newcommand{\DRbar}{\ensuremath{\overline{\text{DR}}}}
\newcommand{\gammaFDH}[2]{\bar\gamma^{#1}_{#2}}
\newcommand{\gammaDR}[2]{\bar\gamma^{#1,\overline{\text{DR}}}_{#2}}
\newcommand{\betaeDR}[1]{\bar\beta^{e,\overline{\text{DR}}}_{#1}}

\begin{document}

\section{Introduction}
Precision calculations in QCD, the electroweak Standard Model, or
supersymmetric models require regularization and
renormalization in intermediate steps. Here we discuss two recent
works where progress in the understanding of fundamental issues and
results of practical relevance were obtained. The topic of
\cite{Sperling} is the renormalization properties of vacuum expecation
values (vevs) in spontaneously broken gauge theories. A method was
developed which leads to a better understanding of the  divergent
renormalization of vevs and their gauge and renormalization-scale
dependence. As a practical result the two-loop renormalization group
$\beta$-functions of vevs in general and supersymmetric gauge theories
were obtained and made available for use in spectrum generators and
explicit calculations.

In \cite{Gnendiger} the structure of infrared divergences was
investigated in different regularization schemes. The recent progress
on computations using unitarity-inspired methods or four-dimensional 
approaches highlights the importance of studying QCD amplitudes in
regularization schemes which differ from conventional dimensional
regularization (CDR). Ref.\ \cite{Gnendiger} studies the relation
between CDR and the four-dimensional helicity (FDH) scheme. The
central results are the proof that the infrared structure can be
predicted in both schemes in a similar way, and that there are simple
translation rules which convert an FDH-regularized amplitude into a
CDR-regularized one. These results are similar to results of
Ref.\ \cite{Kilgore}, and they can be viewed as a continuation of
Refs.\ \cite{Signer}, where a systematic one-loop comparison of the CDR, HV,
FDH, and DRED regularization schemes was carried out and an earlier
factorization problem of DRED was resolved.

\section{Vacuum expectation values and their renormalization}

The renormalization of a scalar field $\phi$ and associated vev $v$
can generically be written as
\begin{align}
\phi+v & \to
\sqrt{Z}\phi+\sqrt{Z}\sqrt{\hat{Z}}v.
\end{align}
Here ${Z}$ is the usual field renormalization constant, and
${\hat{Z}}$ is an additional renormalization constant, which
characterizes to what extent $v$ renormalizes differently from
$\phi$. The field renormalization leads to an anomalous dimension
$\gamma$ of the scalar field, and one can define a similar quantity
$\hat\gamma$, which is defined via $\hat{Z}$ in the same way as
$\gamma$ is defined via $Z$. The renormalization group $\beta$
function for the running vev in the \MSbar\ or \DRbar\ scheme is then
given as 
\begin{align}
\label{betav}
\beta_v &=
\left(\gamma+\hat\gamma\right)v .
\end{align}

A surprising observation was made in the literature in
Refs.\ \cite{TanbetaMSSM}. In the minimal
supersymmetric standard model (MSSM), there are two Higgs doublets $H_{u}$,
$H_{d}$, and there is a cancellation between the two additional
renormalization constants, i.e.\
$\delta \hat{Z}_{H_u}-\delta\hat{Z}_{H_d}=$finite at the one-loop
level. The usual field renormalization constants $\delta Z_{H_u}$,
$\delta Z_{H_d}$ don't share this property. The implication is that
the $\beta$ function for the parameter $\tan\beta=v_u/v_d$ can simply
be written in terms of the usual anomalous dimensions, i.e.
\begin{align}
\delta \hat{Z}_{H_u}-\delta\hat{Z}_{H_d}&=\text{finite}&\Rightarrow&&
\beta_{\tan\beta}&=\left(\gamma_{H_u}-\gamma_{H_d}\right)\tan\beta
\label{tanbetaobservation}
\end{align}
at the one-loop level, with no $\hat\gamma$ contributions.

Furthermore, one may ask why the additional renormalization constant
$\hat{Z}$ is necessary in the first place. Clearly, in non-gauge
theories, i.e.\ theories with only a global symmetry, it is not
required. The point is that local gauge theories can only be quantized
using some kind of gauge fixing. Depending on which gauge fixing is
chosen, the additional renormalization constant $\hat{Z}$ can be
necessary. The method employed in Refs.\ \cite{Sperling} aims to make
this explicit. In the process it explains the observation mentioned
above, and it allows the explicit computation of $\beta_v$ at the
two-loop level in generic gauge theories.

The essence of the method, introduced in
Refs.\ \cite{Kraus} for a slightly different purpose, is to replace
the vev by a background 
(classical) field $\hat\phi$, carry out the renormalization process
(in particular the study of Slavnov-Taylor identities and their
implications on the required independent renormalization constants) in
presence of these background fields, and specialize to $\hat\phi=
v=$const only at the end. Using the background fields, the usual
$R_\xi$ gauge fixing term for the abelian Higgs model can be written
as
  \begin{align}
    F &= \partial^\mu A_\mu +ie\xi(\hat\phi^\dagger\phi - \phi^\dagger \hat\phi).
  \end{align}
As long as $\hat\phi$ is treated as a background field which
transforms covariantly under global gauge transformations, this gauge
fixing term breaks only local, but not global gauge invariance. As a
consequence only renormalization constants in agreement with global
gauge invariance can be necessary. $\hat{Z}$ can appear as the field
renormalization of $\hat\phi$. The main trick is to define a
BRS transformation of the background field, as
\begin{align}
s\hat\phi&=\hat{q},& s\hat{q}&=0.
\end{align}
Then the Slavnov-Taylor identity provides a useful relation for
$\hat{Z}$: this renormalization constant can be directly determined
from the Green function $\Gamma_{\hat{q}K_{\phi}}$, because the
counterterm contribution to this Green function is
$\Gamma^{\text{ct}}_{\hat{q}K_\phi}=-\frac12\delta\hat{Z}$. Here
$K_\phi$ is the source for the BRS transformation of $\phi$, so this
Green function  corresponds to the coupling of the composite operator
$(s\phi)$ to the background field $\hat{q}$. It is an unphysical Green
function but a very useful technical tool. 

\begin{figure}
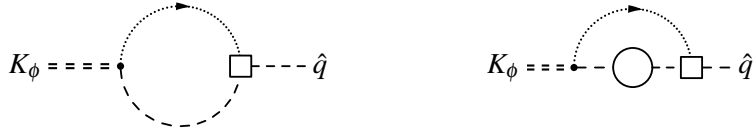

\null\hfill
\begin{picture}(100,30)(0,20)
\put(20,0){\epsfbox{FeynmanRules.5}}
\put(5,35){$K_{\phi}$}
\put(120,35){$\hat q$}
\end{picture}
\hfill
\begin{picture}(120,30)(0,20)
\put(30,7){\epsfbox{FeynmanRules.20}}
\put(15,35){$K_{\phi}$}
\put(110,35){$\hat q$}
\end{picture}
\hfill\null
\caption{Left: The single one-loop diagram to
  $\Gamma_{\hat{q}K_{\phi}}$. The dotted 
  line is a Faddeev-Popov ghost propagator; the dashed line a scalar
  field propagator. Right: Sample two-loop diagram with a fermion loop
  (solid line), which leads to
  corrections involving Yukawa couplings. \label{figure}}
\end{figure}

Fig.\ \ref{figure} (left) shows
the single one-loop diagram in a generic gauge theory.
The only vertex of $K_\phi$ follows from the structure of the BRS
transformation of scalar fields: $K_\phi$ couples to one scalar field
and one Faddeev-Popov ghost; the coupling is the gauge coupling. The
only vertex of $\hat{q}$ originates in the gauge fixing: $\hat{q}$
couples to the prefactor of $\hat\phi$ in the gauge fixing, i.e.\ the
Feynman rule is proportional to $\xi$ and to the gauge coupling. 
Hence the one-loop contribution to this Green function (and thus to
$\delta\hat{Z}$) is proportional only to the gauge parameter $\xi$ and
to the squared gauge coupling of the scalar field. In a generic gauge
theory the \MSbar\ result can be written as
  \begin{align}
 \delta_{a b} \delta \hat{Z}^{(1)}(a) = \frac{1}{(4 \pi)^2} 2 g^2 \xi 
C^2_{ab}(S) \cdot \frac{1}{\epsilon} ,
\label{eqn_Z_hat_CT}
  \end{align}
where $C^2_{ab}(S)=T^A_{ac}T^A_{cb}$ with the generators $T^A$ acting on the scalar fields
$\phi_a$. Hence the result is proportional to $\xi$ and to the squared
gauge couplings of the scalar field. Since all two-loop diagrams must
involve the same two vertices as the ones in Fig.\ \ref{figure} (left), this
proportionality extends to the two-loop result; Fig.\ \ref{figure}
(right) shows a sample two-loop diagram.

This result already explains two statements made earlier. First we see
that in Landau gauge, where $\xi=0$, we obtain $\delta\hat{Z}=0$. This
reflects the fact that in a gauge which does not break global gauge
invariance, the additional renormalization constant $\hat{Z}$ is not
needed. Second, for $\xi\ne0$, $\hat{Z}$ is needed but is proportional to
squared gauge couplings. This explains the observation
(\ref{tanbetaobservation}) since the squared SU(2) and U(1)
gauge couplings of the two Higgs doublets in the MSSM are equal.

\newcommand{\CasiScaGauge}[2]{C_{#1 #2}^{2}(\mathrm{S})}
\newcommand{\DykinScaGauge}[2]{S_{2}^{#1 #2}(\mathrm{S})}
\newcommand{\CasiAdjGauge}[2]{C_{2}^{#1 #2}(\mathrm{G})}
The full two-loop result for $\delta\hat{Z}$, $\hat\gamma$ and
$\beta_v$ has been obtained in Refs.\ \cite{Sperling}. Here we provide
the result for $\gamma$ and $\hat\gamma$ for general supersymmetric
gauge theories:
  \begin{align}
  \gamma_{ab}^{(1)}(\mathrm{S}) \Big|_{\mathrm{SUSY}}^{\text{\DRbar}} &=
\frac{1}{(4 \pi)^2} \left[ g^2 \left(1-\xi \right) \CasiScaGauge{a}{b} -
\frac{1}{2} Y_{apq}^{*} Y_{bpq}^{\phantom{*}} \right]   \; , %
  \label{eqn:gamma_1-Loop_SUSY-complex}\\ 
  \hat{\gamma}_{ab}^{(1)}(\mathrm{S}) \Big|_{\mathrm{SUSY}}^{\text{\DRbar}}
&= \frac{1}{(4 \pi)^2} 2 g^2 \xi \xi' \CasiScaGauge{a}{b}   \; ,%
\label{eqn:hat-gamma_1-Loop-SUSY-complex} \\
\gamma_{ab}^{(2)}(\mathrm{S})\Big|_{\mathrm{SUSY}}^{\text{\DRbar}} &=
\frac{1}{(4 \pi)^4} \bigg\{
   g^4 \left[ \left( \frac{9}{4} -\frac{5}{3} \xi -\frac{1}{4}  \xi^2
\right) \CasiAdjGauge{}{}  - \DykinScaGauge{}{}
\right] \CasiScaGauge{a}{b} %
\label{eqn:gamma_2-loop_DRbar-complex} \\*
&\phantom{= \frac{1}{(4 \pi)^4} \bigg\{ } -2 g^4 \CasiScaGauge{a}{c}
\CasiScaGauge{c}{b}
+ \frac{1}{2}  Y_{arc}^{*} Y_{rpq}^{\phantom{*}} Y_{pqd}^{*}
Y_{bcd}^{\phantom{*}}
  \nonumber \\*
&\phantom{= \frac{1}{(4 \pi)^4} \bigg\{ }   
+ g^2 \left[ \CasiScaGauge{a}{c} Y_{cpq}^{*} Y_{bpq}^{\phantom{*}} -2
  Y_{apq}^{*} \CasiScaGauge{p}{r} Y_{brq}^{\phantom{*}} \right]
 \bigg\}
\nonumber \; ,\\
\hat{\gamma}_{ab}^{(2)}(\mathrm{S})
\Big|_{\mathrm{SUSY}}^{\text{\DRbar/\MSbar}} &=
\frac{\xi \xi'}{(4 \pi)^4} \Bigg\{ g^4   \left[ \frac{7-\xi}{2}
\CasiAdjGauge{}{} \CasiScaGauge{a}{b} -2\left(1 - \xi \right)
\CasiScaGauge{a}{c} \CasiScaGauge{c}{b} \right] %
\label{eqn:hat-gamma_2-Loop-SUSY-complex} \\*
 &\phantom{=\frac{\xi \xi'}{(4 \pi)^4} \Bigg\{ g^4   \Bigg[-2\left(1 - \xi
\right) \CasiScaGauge{a}{a} }- g^2 \CasiScaGauge{a}{c} Y_{cpq}^{*}
Y_{bpq}^{\phantom{*}}  \Bigg\}  \nonumber  \; .
\end{align}
Here $Y_{abc}$ are conventionally normalized superpotential
couplings. To our knowledge, the scalar field anomalous dimension
$\gamma^{(2)}$ has not been provided in the literature before. We
remark that this anomalous dimension is the one of the component
scalar field in Wess-Zumino gauge, and it is not equal to the
superfield anomalous dimension in a supersymmetric gauge fixing. From
these results eq.\ (\ref{betav}) can be used to obtain the $\beta$
function for vevs and for related quantities such as $\tan\beta$.
Refs.\ \cite{Sperling} provide explicit results for models of
practical and phenomenological interest, such as the MSSM, the NMSSM,
and the E$_6$SSM. The general results have been implemented in general
programs such as Sarah \cite{Sarah} and FlexibleSUSY
\cite{FlexibleSUSY}.

\section{Infrared structure of QCD amplitudes in the FDH scheme}

In recent years, new methods have been developed to compute gauge
theory amplitudes, and the understanding of the infrared structure of
these amplitudes has significantly improved. In particular,
Refs.\ \cite{BNGM} have given a prediction for the infrared
$1/\epsilon$ poles in conventional dimensional reduction for arbitrary
QCD amplitudes. Specialized to form factors, this prediction reads
\begin{align}
\label{lnZ}
 \text{ln}\,\mathbf{Z}=
 \left(\frac{\alpha_s}{4\pi}\right)\left(\frac{\Gamma'_1}{4\epsilon^2}+\frac{\Gamma_1}{2\epsilon}\right)+
 \left(\frac{\alpha_s}{4\pi}\right)^2\left(-\frac{3\beta_{20}\Gamma'_1}{16\epsilon^3}+
 \frac{\Gamma'_2-4\beta_{20}\Gamma_1}{16\epsilon^2}+\frac{\Gamma_2}{4\epsilon}\right)+
\ldots
\end{align}
Here $\Gamma'_m = -\,2\,\gamma^{\text{cusp}}_m\, C_{q/g}$,
 $\Gamma_m  =+\,2\,\gamma^{i}_m.$ with the cusp
anomalous dimension $\gamma^\text{cusp}$ and parton anomalous
dimension $\gamma^{i}$; the index $i$ refers to either quark or gluon,
and the index $m$ refers to the coefficient of the respective quantity
of $(\alpha_s/4\pi)^m$; $\beta_{20}$ is the
$(\alpha_s/4\pi)^2$-coefficient of the $\beta$ function for
$\alpha_s$. 

In view of the new computational methods based on unitarity, helicity
and 4-dimensional algebra it is of high interest to study how the
infrared structure depends on the regularization scheme, in particular
how it is modified in schemes such as the four-dimensional helicity
scheme (FDH) and dimensional reduction (DRED). An independent reason
to study these regularization schemes is supersymmetry, which is
broken in CDR, while FDH and DRED preserve supersymmetry to a large
extent. In Refs.\ \cite{Signer} the differences between these schemes
were clarified and it was shown that all these schemes are consistent
regularization schemes. In particular one-loop results of the earlier
literature were shown to be consistent among each other and consistent
with infrared factorization, if the schemes are used appropriately.
The necessity to treat renormalization in the FDH scheme in the way
advocated in \cite{Signer} was also reiterated in
Ref.\ \cite{KilgoreUnitarity}.

Ref.\ \cite{Gnendiger} can be viewed as a
continuation of Refs.\ \cite{Signer} to the two-loop level and as an
extension of Eq.\ (\ref{lnZ}) to other regularization schemes.
Extending the infrared prediction (\ref{lnZ}) to FDH and DRED is
possible by  taking into
account the main insight of Ref.\ \cite{Signer}, which is that FDH and
DRED should be viewed as dimensional regularization with a new type of
parton, a scalar field with multiplicity $N_\epsilon=2\epsilon$, the
$\epsilon$-scalars. The $\epsilon$-scalars have couplings and
anomalous dimensions which differ from the ones of the gluons.
At the two-loop level the only new coupling appearing in the infrared
prediction is $\alpha_e$, the coupling of $\epsilon$-scalars to
quarks. The FDH result corresponding to Eq.\ (\ref{lnZ}) is
\def\FDH{{\scshape fdh}}
\def\DR{{\scshape dr}}
\def\GammaAD{{\Gamma}}
\newcommand{\betaMS}[1]{\bar\beta^{\phantom{e}}_{#1}}
\newcommand{\betaeMS}[1]{\bar\beta^e_{#1}}
\begin{align}
 \text{ln}\,\bar{\mathbf{Z}}&=
 \left(\frac{\alpha_s}{4\pi}\right)\left(\frac{\bar\Gamma'_{10}}{4\epsilon^2}+\frac{\bar\GammaAD_{10}}{2\epsilon}\right)+
 \left(\frac{\alpha_e}{4\pi}\right)\left(\frac{\bar\Gamma'_{01}}{4\epsilon^2}+\frac{\bar\GammaAD_{01}}{2\epsilon}\right)\nonumber\\
 &\quad+
 \left(\frac{\alpha_s}{4\pi}\right)^2
 \left(-\frac{3\betaMS{20}\bar\Gamma'_{10}}{16\epsilon^3}+
 \frac{\bar\Gamma'_{20}-4\betaMS{20}\bar\GammaAD_{10}}{16\epsilon^2}+\frac{\bar\GammaAD_{20}}{4\epsilon}\right)\nonumber\\
 &\quad+
 \left(\frac{\alpha_s}{4\pi}\right)\left(\frac{\alpha_e}{4\pi}\right)
 \left(-\frac{3\betaeMS{11}\bar\Gamma'_{01}}{16\epsilon^3}+
 \frac{\bar\Gamma'_{11}-4\betaeMS{11}\bar\GammaAD_{01}}{16\epsilon^2}+\frac{\bar\GammaAD_{11}}{4\epsilon}\right)\nonumber\\
  &\quad+
 \left(\frac{\alpha_e}{4\pi}\right)^2
 \left(-\frac{3\betaeMS{02}\bar\Gamma'_{01}}{16\epsilon^3}+
 \frac{\bar\Gamma'_{02}-4\betaeMS{02}\bar\GammaAD_{01}}{16\epsilon^2}+\frac{\bar\GammaAD_{02}}{4\epsilon}\right)+
 \mathcal{O}(\alpha^3).
 \label{eq:lnZfinalDRED}
\end{align}
Here the bars denote quantities defined in the FDH scheme, $\beta^e$
is the $\beta$ function for $\alpha_e$, and 
$\Gamma'_{ij}$, $\Gamma_{ij}$, $\beta_{ij}$ and $\beta^e_{ij}$ are the
coefficients of $(\alpha_s/4\pi)^i(\alpha_e/4\pi)^j$ of the respective
quantities. 

The FDH formula differs in two respects from the CDR one. 
First, there are new structures, involving the new $\beta$ function
$\beta^e$ and involving coefficients of the order $\alpha_e$. Second,
all quantities are defined in the FDH scheme and thus contain
contributions from $\epsilon$-scalars and differ at the order
$N_\epsilon$ from the respective CDR quantities.
Ref.\ \cite{Kilgore} has obtained an equivalent formula using a 
slightly different approach.

Concrete two-loop computations of form factors serve to test the
prediction (\ref{eq:lnZfinalDRED}) and to determine the unknown
coefficients. We have computed the two-loop quark and gluon form
factors. In this way $\gamma^{\text{cusp}}$, $\gamma^q$ and $\gamma^g$
are determined at the two-loop level, and there are non-trivial checks
since the system is overconstrained.

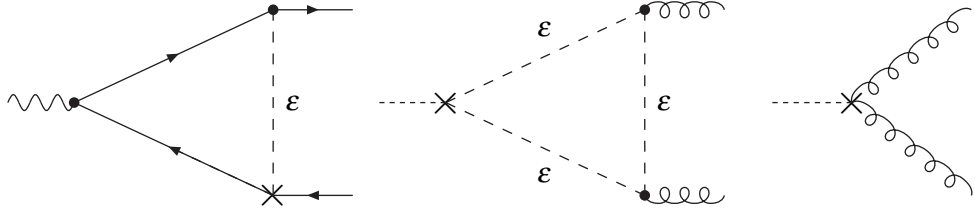
\begin{figure}[t]
\begin{center}
\begin{tabular}{ll}
\scalebox{1}{
\begin{picture}(135,90)(-5,-10)
\Vertex(25,45){2}
\Vertex(100,80){2}
\Text(100,10)[c]{\scalebox{1.5}{{${\times}$}}}
\Photon(0,45)(25,45){3}{3}
\ArrowLine(100,10)(25,45)
\ArrowLine(25,45)(100,80)
\DashLine(62.5,27.5)(100,10){4}
\DashLine(100,10)(100,80){4}
\ArrowLine(100,80)(130,80)
\ArrowLine(130,10)(100,10)
\Text(105,45)[l]{$\epsilon$}
\end{picture}
}
\begin{picture}(135,90)(-5,-10)
\Text(25,45)[c]{\scalebox{1.5}{$\times$}}
\Vertex(100,80){2}
\Vertex(100,10){2}
\DashLine(0,45)(25,45){2}
\DashLine(100,10)(25,45){4}
\DashLine(25,45)(100,80){4}
\DashLine(100,10)(100,80){4}
\Gluon(100,80)(130,80){3}{3}
\Gluon(130,10)(100,10){3}{3}
\Text(62.5,70)[b]{$\epsilon$}
\Text(62.5,15)[b]{$\epsilon$}
\Text(105,45)[l]{$\epsilon$}
\end{picture}
\quad
\begin{picture}(70,90)(-5,-10)
\Text(30,45)[c]{\scalebox{1.5}{$\times$}}
\DashLine(0,45)(30,45){2}
\Gluon(30,45)(75,80){3}{5}
\Gluon(30,45)(75,10){3}{5}
\end{picture}
\end{tabular}
\caption{\label{fig:tlcases}
Sample counterterm diagrams contributing to the quark and gluon form
factors in the FDH scheme. The crosses denote counterterm insertions
involving $\delta \bar Z_{\alpha_e}^{(1)}$, $\delta \bar
Z_{\lambda_\epsilon}^{(1)}$, and $\delta \bar Z_{\lambda}^{(2)}$, respectively.
}
\end{center}
\end{figure}
In the actual calculation of the two-loop form factors in the FDH
scheme, the correct renormalization is particularly important and
non-trivial. We highlight the following two points.
\begin{enumerate}
\item {\em Independent couplings of $\epsilon$-scalars}.
The couplings of $\epsilon$-scalars must be treated as independent
from the respective couplings of gluons. Fig.\ \ref{fig:tlcases} shows
three sample counterterm diagrams involving the renormalization
constants
$\delta \bar Z_{\alpha_e}^{(1)}$, $\delta \bar
Z_{\lambda_\epsilon}^{(1)}$, and $\delta \bar Z_{\lambda}^{(2)}$,
where the upper index denotes the loop order. $\lambda$ is the
effective coupling of the Higgs boson to gluons, $\lambda_\epsilon$ is
the corresponding coupling to $\epsilon$-scalars. The  renormalization
of $\lambda$ also appears in CDR calculations, and here it can be
determined in the same way, although the FDH and CDR results
differ. The renormalization of $\alpha_e$ is known in the literature,
but the renormalization of $\lambda_\epsilon$ is new. The
\MSbar\ value of $\delta \bar{Z}_{\lambda_\epsilon}$ can be
obtained by an explicit one-loop off-shell calculation of the
Higgs--$\epsilon$--$\epsilon$ three-point function.
\item {\em Renormalization schemes \MSbar\ versus \DRbar}.
If CDR is used, the \MSbar\ scheme is simply defined by modified
minimal subtraction of UV $1/\epsilon$ poles. Even in the FDH or DRED
regularization schemes, it is possible to choose
\MSbar\ renormalization, implying that renormalized couplings have the
same meaning as in CDR-based \MSbar. This \MSbar\ scheme is a natural
and advantageous scheme also in the context of FDH and DRED if these
regularizations are viewed as advocated here: As long as the
$\epsilon$-scalars are treated as unrelated to gluons but as some new
scalar fields which happen to have multiplicity $N_\epsilon$, the
\MSbar\ renormalization scheme simply amounts to minimally subtracting
all UV $1/\epsilon$ poles, {\em including} the ones of the form
$N_\epsilon/\epsilon$ from $\epsilon$-scalar loops. In this scheme,
$\beta$ functions and the $\gamma$s involve explicit terms of the
order $N_\epsilon$. 

It is also of interest to study the case of
\DRbar\ renormalization. This amounts to setting
$N_\epsilon=2\epsilon$ and only then minimally subtracting the
remaining $1/\epsilon$ poles. Renormalized couplings in this scheme
differ by finite shifts from the \MSbar\ couplings.
\end{enumerate}
After consistently renormalizing the form factors either in the
\MSbar\ or \DRbar\ schemes, we have shown that in both cases the
infrared singularities are correctly described by
Eq.\ (\ref{eq:lnZfinalDRED}). The respective $\beta$ and $\gamma$
coefficients differ: in the \MSbar\ case, these coefficients  involve terms of the
order $N_\epsilon$; in the \DRbar\ case, no such terms appear, but the
coefficients differ by $\epsilon$-independent terms. As an
illustration, we quote here the results for the quark anomalous
dimension. The CDR (plus \MSbar\ renormalization) results are
\begin{align}
 \gamma_{10}^{q} & = -3\,C_{F},\\
 \gamma_{20}^{q} & = C_{A} C_{F}\left(-\frac{961}{54}-\frac{11}{6}\pi^2+26\zeta(3)\right)
   +C_{F}^{2}\left(-\frac{3}{2}+2\pi^2-24\zeta(3)\right)
\nonumber\\
 & \quad
   +C_{F} N_F\left(\frac{65}{27}+\frac{\pi^2}{3}\right),
\end{align}
the results for FDH plus \MSbar\ 
renormalization or \DRbar\ renormalization differ by ${\cal
  O}(N_\epsilon)$ or by finite terms, respectively,
\begin{align}
 & \gammaFDH{q}{10} = \gamma_{10}^{q},
&& \gammaDR{q}{10} =  \gamma_{10}^{q},
\\
 & \gammaFDH{q}{01} = N_\epsilon \frac{C_F}{2},
&& \gammaDR{q}{01} = 0,\\
 & \gammaFDH{q}{20} = \gamma_{20}^{q} + N_\epsilon \Big(\frac{167}{108}+\frac{\pi^2}{12}\Big)C_A C_F,
&& \gammaDR{q}{20}  = \gamma_{20}^{q} + \frac{17}{9}\,C_A C_F,
\\
 & \gammaFDH{q}{11} = N_\epsilon \Big[\frac{11}{2}C_A C_F-\Big(2+\frac{\pi^2}{3}\Big)C_F^2\Big],
& & \gammaDR{q}{11} = -\,\betaeDR{11}\,C_F,\\
 &\gammaFDH{q}{02}  = - N_\epsilon \frac{3}{4} C_F N_F - N_\epsilon^2\frac{C_F^2}{8},
& & \gammaDR{q}{02} = -\,\betaeDR{02}\,C_F ,
\end{align}
with the non-vanishing $\beta$-coefficients
\begin{align}
 \betaeDR{11}&=\left.\betaeMS{11}\right|_{N_\epsilon=0}
 =6\,C_F,\phantom{\frac{11}{3}}
 \label{eq:betaeDR11}\\*
 \betaeDR{02}&=\left.\betaeMS{02}\right|_{N_\epsilon=0}
 =-4\,C_F+2\,C_A-N_F.
 \phantom{\frac{1}{1}}
 \label{eq:betaeDR02}
\end{align}
The \DRbar\ results are relevant since they answer a question of the
``Supersymmetry Parameter Analysis'' report \cite{SPA}: the
\DRbar\ scheme is a consistent scheme not only for UV \cite{DS05} but
also for infrared divergences at the multi-loop level.

The \MSbar\ result has the important application of transition rules
between FDH and CDR regularized amplitudes. For the example of the
quark form factor $F_q$ we can define the combination $Q^{(2)}\equiv
 F^{2l}_q-\frac{1}{2}\left(F^{1l}_q\right)^2$ and obtain
\begin{align}
\left[{Q}^{(2)}- \text{ln}\,\mathbf{\bar Z}_q^{(2)}\right]^{\text{FDH}}=
\left[{Q}^{(2)}- \text{ln}\,\mathbf{Z}_q^{(2)}\right]^{\text{CDR}}+{\cal
  O}(N_\epsilon \epsilon^0),
\end{align}
which allows to translate e.g.\ an FDH-amplitude into a CDR-amplitude or
vice versa.

\end{document}